\documentclass[aps,pre,twocolumn,superscriptaddress,floatfix,longbibliography]{revtex4-2}
\usepackage{graphicx}
\usepackage{amsmath}
\usepackage{amssymb}
\usepackage{bm}
\usepackage{dcolumn}
\usepackage[colorlinks=true,linkcolor=blue,citecolor=blue,urlcolor=blue]{hyperref}

\begin{document}

\title{High-resolution mapping of the phase boundary of the honeycomb-lattice Blume-Capel ferromagnet}

\author{Eluiz C.M. Filho}
\affiliation{Instituto Federal de Goi\'as, Rua 76, Centro, 74055-110 Goi\^ania, GO, Brazil}

\author{Claudio J. DaSilva}
\affiliation{Instituto Federal de Goi\'as, Rua 76, Centro, 74055-110 Goi\^ania, GO, Brazil}
\affiliation{Instituto de Física, Universidade Federal de Goiás, C.P. 131 CEP 74001-970, Goiânia, Goiás, Brazil.}

\date{\today}

\begin{abstract}
We present a high-resolution numerical study of the spin-1 Blume-Capel ferromagnet on the honeycomb lattice ($z=3$). We use the Wang-Landau algorithm to estimate the joint density of states $\Omega(E_1,E_2)$ and combine it with the field-mixing scaling method to map the first-order transition line and locate the tricritical point at $(\Delta_t,T_t)=[1.4820(2),\,0.4022(7)]$. Histogram reweighting of Metropolis simulations for system sizes $L=36$ and $48$ provides an independent consistency check, revealing a crossing temperature near $0.4039$, which is compatible with the Wang-Landau extrapolation. From the field-mixing analysis,  we obtain the mixing parameter $r=-1.585(19)$. Additionally, the slope of the multicanonical coexistence line yields a more precise estimate of $r=-1.587(5)$. When normalized by the zero-temperature field $\Delta_0 = zJ/2$, the estimates for honeycomb, square, and triangular lattices yield values of $r/\Delta_0$ between $-1.057$ and $-1.067$, which corresponds to a magnitude that is about $6\%$ larger than the mean-field value of $-1$. For the simple-cubic lattice, literature values for transition fields give $r/\Delta_0 = -1.04(2)$, situating it between the 2D values and the mean field. However, its uncertainty, primarily stemming from extrapolation to the tricritical temperature, does not allow for a clear resolution of dimensionality dependence. Finite-size scaling of the width $\sigma_Q$ of the order-parameter distribution gives $y_t=1.795(7)$, which is consistent with the tricritical Ising value $9/5$. The scaling of the fourth-order cumulant is compatible with $y_g=4/5$. Deep within the first-order regime, the cumulant approaches the expected two-phase limit, with $U_Q\to2/3$. Multicanonical simulations up to $L=64$, utilizing weights constructed directly from the Wang-Landau density of states for $L\le24$ and rescaled and refined for larger systems, yield an interface tension exponent $\mu=1.265(26)$, which is consistent with the tricritical prediction $(d-1)/y_g=5/4$. Finally, we map the continuous branch of the phase boundary from the same density-of-states data, obtaining $T_c(\Delta=0)=1.157$; estimates sufficiently close to the tricritical point are affected by crossover from Ising to tricritical scaling.
\end{abstract}

\maketitle

\section{Introduction}

The Blume-Capel (BC) model~\cite{Blume1966,Capel1966} is a spin-1 Ising model that incorporates a single-ion crystal-field anisotropy. Its phase diagram features a line of continuous transitions and a line of first-order transitions that separate ordered and disordered phases; these two lines intersect at a tricritical point. This topology has established the BC model as a fundamental framework for investigating tricriticality, metamagnets, and finite-size scaling~\cite{DombLebowitz}. In zero magnetic field, the Hamiltonian is given by 
\begin{align} 
\mathcal{H} = -J\sum_{\langle ij\rangle}\sigma_i\sigma_j + \Delta\sum_i \sigma_i^2 = J E_1 + \Delta E_2, 
\label{eq:ham} 
\end{align} 
with $\sigma_i \in \{-1,0,+1\}$ and $J>0$. Energies are measured in units of $J$, and temperatures in units of $J/k_B$. In this context, $N=L\times L$ represents the number of sites in our finite-lattice setup, and $E_2$ is the number of nonzero spins. At $T=0$, the ferromagnetic state is favored as long as its energy gain per spin, $zJ/2$, exceeds the crystal-field penalty. This condition places the specific point $(\Delta_0 = zJ/2,T=0)$ on the phase boundary. The BC model also provides an intuitive interpretation within the framework of lattice gases~\cite{Ibagon2021}. In this viewpoint, the states $\sigma_i=\pm1$ represent occupied sites, whereas $\sigma_i=0$ represents a vacancy or an inert component. With the sign convention of Eq.~\eqref{eq:ham}, $-\Delta$ plays the role of an effective lattice-gas chemical potential. This conceptual structure is similarly connected to the Blume-Emery-Griffiths (BEG) model, which has been used to examine phase separation and the \textit{lambda} transition in mixtures of helium isotopes ($^3$He-$^4$He)~\cite{Blume1971}.

Comparing different lattice geometries is valuable for distinguishing universal critical properties from non-universal quantities that are influenced by microscopic connectivity and geometry. The BC model has been investigated on several lattices, including the square~\cite{Silva2006,Kwak2015,Zierenberg2017} and triangular lattices~\cite{Mataragkas2025}, as well as on different three-dimensional lattices~\cite{Mataragkas2026}. Among regular two-dimensional lattices, the honeycomb lattice is noteworthy because it is bipartite and has a coordination number of $z=3$, the lowest among these lattices. This low coordination makes the honeycomb lattice particularly useful for examining how connectivity affects non-universal thermodynamic quantities in the BC model. Since each site has only three nearest neighbors, the energetic contribution of interactions between spins is diminished relative to lattices with higher connectivity, such as the square ($z=4$) and triangular ($z=6$) lattices. In this context, the occupation of the $\sigma_i=0$ state decreases the number of effective magnetic interactions between neighboring spins, which can significantly influence the stability of magnetic order. Consequently, non-universal quantities such as the transition temperatures and the location of the tricritical point depend on lattice structure, even though the related universality class is not expected to depend solely on the coordination number. 

The BC model has been extensively studied across various lattices; however, a high-resolution numerical characterization of the phase boundary in the honeycomb lattice remains, to our knowledge, unavailable. For this geometry, Kaneyoshi obtained results for the spin-1 Ising model using a correlated effective-field treatment, including a characterization of the tricritical region~\cite{Kaneyoshi1990}. Yet, a similarly precise numerical determination of the phase boundary and tricritical region for the honeycomb lattice has not been reported. This gap is particularly relevant because the honeycomb lattice has the lowest coordination number among regular two-dimensional lattices, providing an opportunity to investigate how connectivity influences non-universal parameters associated with the tricritical region and to assess the validity of results obtained using the effective-field treatment. It is also important to note that a Monte Carlo study of the honeycomb BEG model has already highlighted accurate tricritical lines on the honeycomb lattice\cite{Booth1993}, although that model is more general than the zero-biquadratic BC limit.

The characterization of the tricritical region involves not only identifying the point where the first-order and second-order transition lines converge but also determining the scaling properties relevant to the associated universality class. In this context, the field mixing scaling method offers a systematic approach for identifying the appropriate scaling fields and their corresponding operators near the tricritical point~\cite{Wilding1992,Bruce1992,Wilding1996}. In particular, the mixing parameter $s$ is included in the definition of the ordering operator $Q$, while the parameter $r$ establishes its absolute normalization. Finite-size estimates of these parameters, along with tests of tricritical scaling, provide a more comprehensive characterization of the tricritical region and facilitate quantitative comparisons among different lattice geometries.

Monte Carlo methods are key numerical tools used to study phase transitions in spin systems~\cite{Landau2014}. Notable techniques include Metropolis sampling, histogram reweighting (HR), multicanonical methods (MUCA), and entropic sampling based on the Wang-Landau (WL) algorithm~\cite{Wang2001,Wang2001a}. Among these methods, the WL algorithm allows for the direct estimation of a system's density of states, enabling the reconstruction of thermodynamic quantities across various temperatures and control parameters. For the BC model, utilizing a joint density of states in two energy variables allows simultaneous consideration of contributions from both interaction and crystal-field terms, thus providing access to the parameter space $(T,\Delta)$ from a single density of states estimate. This approach is especially effective in determining phase boundaries and exploring the tricritical region. Furthermore, combining WL data with histogram reweighting and multicanonical methods extends the analysis to larger systems~\cite{Mataragkas2025}. In this work, we apply these complementary methods to achieve a high-resolution characterization of the honeycomb-lattice BC phase diagram.

To achieve our goals, we employ entropic simulations based on the WL algorithm to determine the joint density of states $\Omega(E_1, E_2)$. This information allows us to investigate the phase boundary in the thermodynamic parameter space. Our analysis is further enhanced by employing the field mixing method, histogram reweighting, and multicanonical simulations. These techniques enable us to identify the tricritical point, determine the field-mixing parameter, and test tricritical scaling along with the vanishing of the interface tension. Additionally, we map the continuous branch of the phase boundary from the same density of states, which facilitates the construction of the phase diagram across the entire temperature range considered in this study. For the WL calculations, systems with linear dimensions up to $L=24$ are considered, while the combination with the other techniques allows the investigation to be extended to larger systems, reaching $L=64$.

\section{Numerical methods}

\subsection{Wang-Landau joint density of states}
 
We employ a single-walker WL algorithm~\cite{Wang2001,Wang2001a} in the two-dimensional $(E_1,E_2)$ space and estimate $\Omega(E_1,E_2)$ for $L=8,12,16,18,24$. We use the variant of Refs.~\cite{Caparica2012,Caparica2014}: the density of states is updated once per Monte Carlo sweep rather than after every spin-flip attempt; microcanonical averages are accumulated only from the eighth WL level onward, when $\Omega$ is already close to converged; and each run is halted automatically once the temperature of the specific-heat maximum becomes stable, rather than at a fixed final modification factor. The usual flatness criterion, $80\%$ for $H(E_1,E_2)$, was adopted. To save computer time~\cite{Ferreira2018}, all runs for a given $L$ are initialized from a common simulation carried to level $f_6$, after which ten trajectories are evolved independently; this procedure reduces the total computational cost by about $60\%$. Because the WL procedure determines $\Omega$ only up to a multiplicative constant, we normalize each run individually using the exact sum rule
\begin{align}
  \sum_{E_1,E_2}\Omega(E_1,E_2) = 3^N,
  \label{eq:sumrule}
\end{align}
which follows from the fact that each of the $N$ spins has exactly three unconstrained states. The normalized runs are combined only after this step. As a global consistency check, we verified that the $T\to\infty$ limits $n \to 2/3$ and $\langle M^2\rangle \to 2N/3$ are recovered to within $10^{-3}$.
 
Once $\Omega$ is known, the joint probability at any $\beta \equiv 1/T$ and $\mu \equiv \Delta/T$ follows without further
sampling,
\begin{align}
  P(E_1,E_2;\beta,\mu) = \frac{1}{\mathcal{Z}}\,\Omega(E_1,E_2)\, e^{-\beta E_1 - \mu E_2}.
  \label{eq:pjoint}
\end{align}
Observables that are not functions of $(E_1,E_2)$ alone (in particular the magnetization) were accumulated as microcanonical averages $\langle |M|^k \rangle (E_1,E_2)$ during the production stage. The smallest lattice, $L=8$, does not provide a stable field-mixing solution: the discreteness of the $(E_1,E_2)$ grid at $N=64$ is such that the peak positions of the relevant distribution jump between adjacent grid points as the mixing parameter is varied, and no stable solution of the equal-height condition exists. It is excluded from all results below.

\subsection{Multicanonical simulations}
\label{sec:mucamethod}

In the first-order regime, the two coexisting phases are separated by a free-energy barrier that increases linearly with $L$. To measure this barrier, we performed multicanonical (MUCA) simulations~\cite{Berg1992} at fixed temperature, which effectively flattened the marginal distribution of $E_2$. The acceptance probability for single-spin updates is given by:
\begin{align}
  \min\!\left[1,\; e^{-\beta\,\delta E_1}\,\frac{W(E_2')}{W(E_2)}\right],
  \label{eq:mucaacc}
\end{align}
ensuring that $E_1$ remains canonically distributed at each $E_2$ while the weights $W$ compensate the barrier in $E_2$. The histogram of $E_2$ is flat when $W=1/\rho_\beta$, with
\begin{align}
  \rho_\beta(E_2) = \sum_{E_1}\Omega(E_1,E_2)\,e^{-\beta E_1}.
  \label{eq:rhobeta}
\end{align}
For $L\le24$, this sum is obtained directly from the WL density-of-states estimate, allowing us to supply the corresponding MUCA weights without a separate iterative construction phase. For $L=32$, $48$ and $64$, the weights from the next smaller size were extensively rescaled using: $\ln W_L(n)\simeq(L/L')^{d}\ln W_{L'}(n)$ at fixed $n=E_2/N$. These weights were then refined using:
\begin{align}
  \ln W_{\rm new} = \ln W_{\rm old} - \ln H,
  \label{eq:refine}
\end{align}
where $H$ is the histogram of $E_2$ produced with $W_{\rm old}$. This refinement process was iterated with short runs until the condition $\ln(H_{\max}/H_{\min})\le2$ was satisfied.

With a fixed set of MUCA weights, the production histogram yields the reweighted estimate $\rho_\beta=H/W$. From this, the distribution at any crystal field can be obtained through one-dimensional reweighting:
\begin{align}
  P_L(E_2;T,\Delta) \propto \rho_\beta(E_2)\,e^{-\beta\Delta E_2}.
  \label{eq:mucarw}
\end{align}

For each combination of $(L, T)$, we performed ten independent production runs, consisting of approximately $10^6$ sweeps, with up to $2 \times 10^7$ sweeps for $L = 64$.

Statistical uncertainties were estimated based on the run-to-run dispersion. If the $\chi^2$ value from a subsequent fit exceeded the number of degrees of freedom $\nu$, uncertainties were adjusted by $\sqrt{\chi^2/\nu}$. Each dataset underwent two verification checks. First, since $\ln \rho_\beta$ and $H$ are recorded on a bin-by-bin basis, we could reconstruct the weights of each run as $\ln W = \ln H - \ln \rho_\beta$, requiring all runs included in a combination to share the same $W$. Second, reweighting to $\Delta = 0$ was confirmed to reproduce the nearly saturated ferromagnet condition, $\langle n \rangle = 1 - O(e^{-zJ/T})$, which was validated to within $10^{-4}$ for every dataset used in the analysis.

\subsection{Histogram reweighting}

To explore sizes beyond those achievable with the two-parameter WL calculation, we conduct conventional Metropolis simulations at fixed parameters $(\beta_0, \mu_0)$ for lattice sizes $L = 36$ and $48$. We accumulate the joint histogram $H(E_1, E_2)$ and apply reweighting according to the following equation~\cite{Ferrenberg1988}:
\begin{align}
  P(E_1,E_2;\beta,\mu) \propto H(E_1,E_2;\beta_0,\mu_0)\,e^{-\beta' E_1 - \mu' E_2},
  \label{eq:hr}
\end{align}
where $\beta' = \beta - \beta_0$ and $\mu' = \mu - \mu_0$. The reference field $\Delta_{\rm ref}$ is selected based on the extrapolation from the smaller lattices to ensure that the solution is positioned near the center of the reweighting window; our analysis utilizes $|\mu'| \lesssim 10^{-3}$. Each simulation run consisted of $2 \times 10^6$ lattice sweeps following a period of equilibration.

\subsection{The field mixing scaling method}

Near the tricritical point, the parameters $\beta$ and $\mu$ are not the relevant scaling fields by themselves; rather, the relevant fields are linear combinations defined as follows~\cite{Wilding1992,Bruce1992}:
\begin{align}
  \lambda &= (\mu-\mu_{t}) + r\,(\beta-\beta_{t}),\\
  g &= (\beta-\beta_{t}) + s\,(\mu-\mu_{t}),
\end{align}
where $s$ and $r$ are size-dependent mixing parameters. The operator conjugate to $\lambda$ is given by 
\begin{align}
  Q = \frac{1}{1-rs}\,(n - s\,\ell),
  \label{eq:Q}
\end{align}
where $\ell \equiv L^{-d}E_1$ and $n \equiv L^{-d}E_2$.

For each temperature, we identify the pair $(s_L, \Delta^*_L)$ that creates a symmetric, double-peaked distribution of the normalized variable $\tilde{Q} \equiv (Q - \langle Q \rangle)/\sigma_Q$. This is achieved by imposing the equal-population condition $\sum_{\tilde{Q}>0} P(\tilde{Q}) = 1/2$ and ensuring both maxima have equal heights. A small Gaussian broadening is applied (with $10^{-2}$ for $L \geq 16$ and $2 \times 10^{-2}$ for smaller $L$) to facilitate the evaluation of peak heights.

It is important to note that the parameter $r$ does not influence the determination of $\Delta^*_L$, $s_L$, or the cumulant $U_Q$ defined below. The parameter $r$ appears in Eq. \eqref{eq:Q} solely through the global factor $(1 - rs)^{-1}$, and all three quantities remain invariant under the transformation $Q \to cQ$. The parameter $r$ only becomes significant in Sec. \ref{sec:exponents}, where we test the absolute scale of \(Q\).

\section{The tricritical point}
\label{sec:tricritical}

\subsection{First-order line}

The pseudo-transition fields, denoted as $\Delta^*_L(T)$, derived from the field-mixing analysis, exhibit the expected behavior for first-order finite-size transitions. This behavior is described by the equation:
\begin{align}
  \Delta^*_L = \Delta^*_\infty + b\,L^{-2},
  \label{eq:fss1st}
\end{align}
with very small deviations. For every temperature studied, the four system sizes $L = 12, 16, 18, 24$ demonstrate a consistent linear dependence on $L^{-2}$, with deviations from the fit remaining below $2 \times 10^{-5}$ on the first-order side and $3\times10^{-5}$ overall. The extrapolated values are summarized in Table~\ref{tab:firstorder}.

It is important to note that the $L^{-2}$ form of Eq.~\eqref{eq:fss1st} represents the first-order finite-size scaling law. Above the tricritical point ($T_{t}$), the appropriate shift exponent changes to $L^{-1/\nu}$, where the Ising value of $\nu$ is $1$. For system sizes of $L \leq 24$, both forms fit the data nearly equally well, with maximum deviations of $3 \times 10^{-5}$ and $7 \times 10^{-5}$ respectively. However, these two approaches differ in their extrapolated values by up to $1.6 \times 10^{-3}$. Therefore, the entries above $T_{t}$ should be considered pseudo-transition points rather than precise determinations of the critical line.

A direct validation of Eq.~\eqref{eq:fss1st} is provided by the histogram-reweighting data. Extrapolating the results from the WL sampling for $L = 12$ to $24$ at $T = 0.400$ predicts $\Delta^*_{36} = 1.48283$. In contrast, the $L=36$ Metropolis run yields a result of $1.48272$, illustrating agreement to one part in $10^{4}$ between two entirely different simulation methodologies.
\begin{table}[tb]
\caption{Pseudo-transition field $\Delta^*_\infty$ along the phase boundary, from the $L^{-2}$ extrapolation of the WL field-mixing data for system sizes ($L=12,16,18,24$). The last column of the table indicates the nature of the transition at each temperature, highlighting the tricritical point at $T_{t} = 0.4022(7)$ that separates the two regimes. The last three rows of the table present results for a single lattice size with $L=24$.}
\label{tab:firstorder}
\begin{ruledtabular}
\begin{tabular}{lcl}
$T$ & \multicolumn{1}{c}{$\Delta^*_\infty$} & Character \\
\hline
0.360 & 1.490660(9)  & first order \\
0.380 & 1.487063(16) & first order \\
0.390 & 1.484923(20) & first order \\
0.395 & 1.483762(22) & first order \\
0.400 & 1.482538(25) & first order \\
\hline
\multicolumn{3}{c}{$T_{t}=0.4022(7)$, $\Delta_{t}=1.4820(2)$} \\
\hline
0.405 & 1.481253(29) & continuous \\
0.410 & 1.479899(31) & continuous \\
0.415 & 1.478482(36) & continuous \\
0.420 & 1.476987(40) & continuous \\
\hline
\multicolumn{3}{c}{$L=24$ only} \\
\hline
0.250 & 1.499517 & first order \\
0.300 & 1.497480 & first order \\
0.350 & 1.492621 & first order \\
\end{tabular}
\end{ruledtabular}
\end{table}

\subsection{Location of the tricritical point}

At tricriticality, the appropriately scaled order-parameter distribution becomes size-independent. We investigate this behavior through the fourth-order cumulant defined as:
\begin{align}
  U_Q^L = 1 - \frac{\langle \tilde{Q}^4\rangle}{3\,\langle \tilde{Q}^2\rangle^2},
  \label{eq:UQ}
\end{align}
which is evaluated along the line $\Delta^*_L(T)$. Curves for different system sizes $L$ intersect at the temperature $T_t$; these crossing temperatures are summarized in Table \ref{tab:crossings}.
\begin{table}[tb]
\caption{Crossing temperatures of the cumulant curves $U_Q^L(T)$ for adjacent pairs of lattice sizes.}
\label{tab:crossings}
\begin{ruledtabular}
\begin{tabular}{lld}
Pair & Method & \multicolumn{1}{c}{$T_{\rm cross}$} \\
\hline
$12\times16$ & WL$\times$WL & 0.39903 \\
$16\times18$ & WL$\times$WL & 0.40024 \\
$18\times24$ & WL$\times$WL & 0.40069 \\
$24\times36$ & WL$\times$HR & 0.40186 \\
$36\times48$ & HR$\times$HR & 0.40394 \\
\end{tabular}
\end{ruledtabular}
\end{table}

Fitting the crossings for the three WL$\times$WL configurations to the form $T_L = T_\infty + a L^{-\omega}$ requires selecting a correction exponent $\omega$, which is not independently determined by our data. Using the triangular-lattice value $\omega = 1.92(51)$ from Ref.~\cite{Mataragkas2025} and propagating its uncertainty, we find the following values for $T_t$: $0.40297$, $0.40221$, and $0.40177$ for $\omega = 1.41$, $1.92$, and $2.43$, respectively. This leads to:
\begin{align}
  T_{t} = 0.4022 \pm 0.0004\,({\rm stat}) \pm 0.0006\,(\omega) = 0.4022(7).
\end{align}

A broader scan across $\omega \in [1.4, 2.4]$ indicates that $T_t$ shifts only between $0.40179$ and $0.40299$, showing that the estimate is stable at the $10^{-3}$ level across this range of correction exponents. We highlight that the primary source of uncertainty comes from this systematic variation, rather than from the statistical error of the fit.
\begin{figure*}[tb]
   \includegraphics[width=\textwidth]{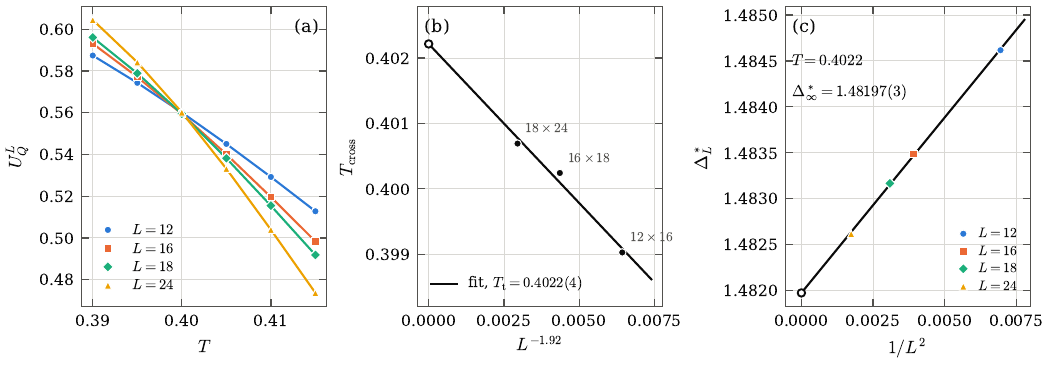}
   \caption{\label{fig:fig1} (a) Cumulant $U_Q^L$ along $\Delta^*_L(T)$. (b) Extrapolation of the crossing temperatures. (c) $L^{-2}$ extrapolation of $\Delta^*_L$ at $T=T_{t}$.}
\end{figure*}

The purely HR crossing at $36\times48$ gives a result of $0.4039(5)$, while the cross-method crossing at $24\times36$ yields $0.4019$. Both results are consistent with the extrapolated value and serve as complementary checks, similar to the HR$(36,48)$ verification from Ref.~\cite{Mataragkas2025}. It is important to note that the uncertainty reported for the $36\times48$ crossing is not a statistical error, as only a single random-number sequence was used at each of these two sizes, which means no run-to-run variance is available. Instead, the reported spread results from varying the interpolation method for $U_Q^L(T)$ among linear, quadratic, and cubic-spline forms, and should be interpreted as a lower bound on the true uncertainty.

Evaluating the $L^{-2}$ extrapolation of $\Delta^*_L$ at $T=T_{t}$ yields the tricritical field
\begin{align}
  \Delta_{t} = 1.4820(2),
\end{align}
where the quoted uncertainty is primarily influenced by the propagated error of $T_t$ rather than the extrapolation itself.

Figure~\ref{fig:fig1} illustrates the three steps involved in determining the tricritical point. Figure~\ref{fig:fig1}(a) displays the cumulant curves $U_Q^L(T)$ evaluated along $\Delta^*_L(T)$; the size-dependent intersections of these curves provide the finite-size crossing temperatures listed in Table~\ref{tab:crossings}. Figure~\ref{fig:fig1}(b) shows the extrapolation of these crossings toward the thermodynamic limit, leading to $T_t = 0.4022(7)$ after accounting for the uncertainty in the correction exponent $\omega$. Finally, Fig.~\ref{fig:fig1}(c) presents the $L^{-2}$ extrapolation of $\Delta_L^*$ at $T = T_t$, yielding $\Delta_t = 1.4820(2)$. Therefore, the coordinates of the tricritical point are obtained through distinct finite-size extrapolations of the cumulant crossings and the pseudo-transition field, respectively.
\begin{figure*}[tb]
   \includegraphics[width=\textwidth]{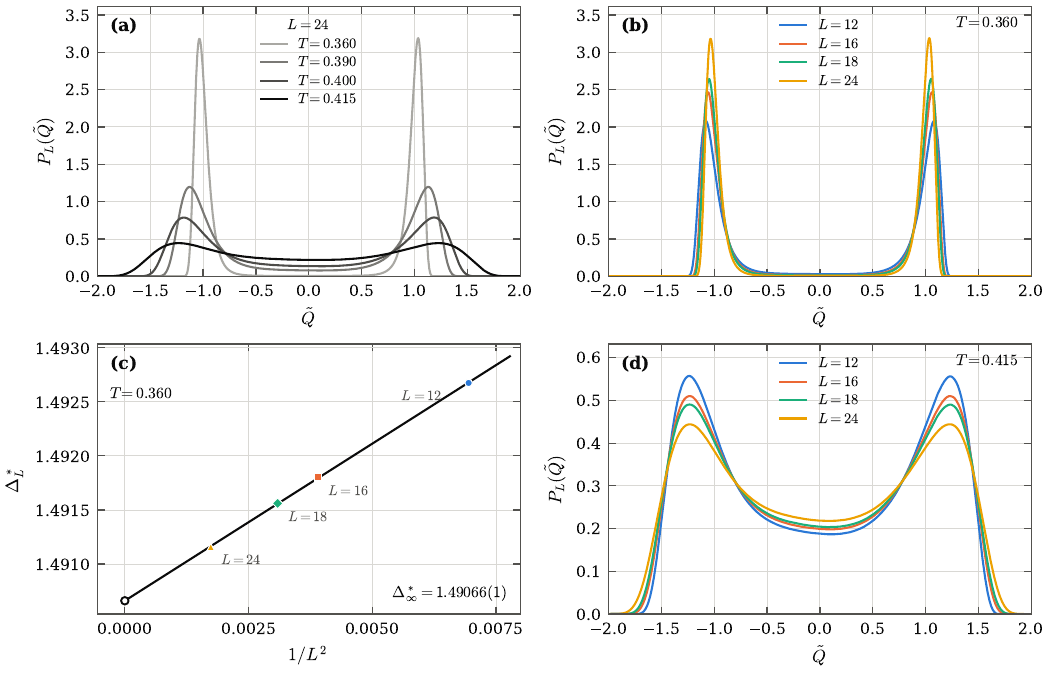}
   \caption{\label{fig:fig2} (a) $P_L(\tilde{Q})$ for $L=24$ along the transition line. (b) Size dependence at $T=0.360$, in the first-order regime: the peaks sharpen with $L$. (c) $L^{-2}$ extrapolation of $\Delta^*_L$ at $T=0.360$. (d) Size dependence at $T=0.415$, above the tricritical point: the peaks flatten with $L$.}
\end{figure*}

Table~\ref{tab:compare} compares the tricritical coordinates and the ratio $\Delta_t / \Delta_0$ with effective-field predictions and results from other two-dimensional lattices. It is worth noting that the ratio $\Delta_t / \Delta_0$ varies only slightly among the different lattice geometries listed.

\subsection{Deep first-order regime}

Below the temperature $T_{t}$, the two coexisting phases become well-defined, and the distribution $P_L(\tilde{Q})$ approaches two delta peaks at $\tilde{Q} = \pm 1$. According to Eq.~\eqref{eq:UQ}, this leads to $U_Q = 1 - \frac{1}{3} = \frac{2}{3}$. This limit is approached smoothly; for $L = 24$, we find $U_Q = 0.666573$, $0.666104$, and $0.661878$ at temperatures $T = 0.25$, $0.30$, and $0.35$, respectively. The deviation from $\frac{2}{3}$ increases monotonically as $T_{t}$ is approached from below. This behavior provides a clear finite-size signature of a first-order transition within this temperature range.
\begin{table}[!h]
\caption{Tricritical point of the honeycomb BC model compared with effective-field results and with other two-dimensional lattices. CEFT and EFT values are from Ref.~\cite{Kaneyoshi1990}, converted to the sign convention of Eq.~\eqref{eq:ham} ($\Delta=-D$). Square- and triangular-lattice values are from Refs.~\cite{Kwak2015} and \cite{Mataragkas2025}.}
\label{tab:compare}
\begin{ruledtabular}
\begin{tabular}{lddd}
 & \multicolumn{1}{c}{$\Delta_{t}$}
 & \multicolumn{1}{c}{$T_{t}$}
 & \multicolumn{1}{c}{$\Delta_{t}/\Delta_0$} \\
\hline
honeycomb, EFT   & 1.4181 & 0.7101 & 0.945 \\
honeycomb, CEFT  & 1.4181 & 0.5865 & 0.945 \\
honeycomb, this work & 1.4820 & 0.4022 & 0.988 \\
square           & 1.9660 & 0.6080 & 0.983 \\
triangular       & 2.9388 & 0.9850 & 0.980 \\
\end{tabular}
\end{ruledtabular}
\end{table}

Figure~\ref{fig:fig2} summarizes the evolution of the mixed-order parameter distribution and the finite-size behavior used to identify the two regimes. Figure~\ref{fig:fig2}(a) displays $P_L(\tilde{Q})$ for $L = 24$ at key points along the transition line, illustrating how the distribution evolves as it approaches and crosses the tricritical region. Figure~\ref{fig:fig2}(b) shows the size dependence at $T = 0.360$, where the two peaks become sharper with increasing $L$ (with peak heights of $2.05$, $2.47$, $2.63$, and $3.19$ for $L = 12, 16, 18,$ and $24$, respectively), which is consistent with first-order transition behavior. Figure~\ref{fig:fig2}(c) presents the corresponding $L^{-2}$ extrapolation of $\Delta_L^*$ at this temperature, exemplifying the finite-size scaling used to determine the first-order coexistence line. In contrast, Fig.~\ref{fig:fig2}(d) illustrates that at $T = 0.415$, the peaks flatten with increasing $L$ (with heights of $0.56$, $0.51$, $0.49$, and $0.44$, indicating behavior characteristic of the continuous-transition side of the tricritical point.

\section{Field-mixing parameter and tricritical exponents}
\label{sec:exponents}

\subsection{The scaling parameter $r$}

The direction of the coexistence curve in the $\beta$-$\mu$ plane determines the second mixing parameter, defined as $r = -\left( \frac{{\rm d}\mu}{{\rm d}\beta} \right)$ evaluated at the tricritical point. Using the extrapolated first-order line from Table~\ref{tab:firstorder}, we obtain:
\begin{align}
  r = -1.585 \pm 0.002\,({\rm stat}) \pm 0.019\,({\rm syst}).
\end{align}
The uncertainty is primarily influenced by the systematic component, which has been estimated from the variability of $r$ under reasonable variations in the analysis. This includes adjusting the fitting window to $|T - T_{t}| \le 0.015$ and $0.030 $, restricting the fit to temperatures below $T_{t}$, and — the most significant factor — altering the extrapolation of $\Delta^*_L$ above $T_{t}$ from the $L^{-2}$ form to the Ising form $L^{-1}$, as detailed in the caption of Table~\ref{tab:firstorder}. The resulting range for $r$ is $-1.615 \le r \le -1.578$.

To facilitate comparison, we normalize $r$ by the zero-temperature field $\Delta_0 = \frac{zJ}{2}$. For the honeycomb lattice, we find $\frac{r}{\Delta_0} = -1.057(13)$. An independent measurement from the slope of the multicanonical coexistence line, presented in Sec.~\ref{sec:muca}, yields $\frac{r}{\Delta_0} = -1.058(3)$. Similarly, applying the same method to the WL joint densities of states generated for the square lattice (with sizes $L = 12, 16, 18, 24$), we obtain $r = -2.129(1)_{\rm stat}$, which translates to $\frac{r}{\Delta_0} = -1.065$.

As an additional verification, we apply the same analysis to the high-precision transition fields tabulated by Kwak et al.~\cite{Kwak2015}, which extend to $L = 48$ and cover 28 temperatures. This results in $r = -2.122(23)$, or $\frac{r}{\Delta_0} = -1.061$. There is an agreement of about 0.3\% between our results for $L \le 24$ and those based on considerably larger systems. The value for the triangular lattice, reported in Ref.~\cite{Mataragkas2025}, is $r = -3.2009(3)$, corresponding to $\frac{r}{\Delta_0} = -1.067$.

To our knowledge, these are the first determinations of $r$ for the honeycomb, square, and simple cubic lattices. Although the field-mixing formalism has been applied to all three, the parameter $r$ has not been determined in previous studies~\cite{Kwak2015, Mataragkas2026}, with the triangular lattice being the only exception~\cite{Mataragkas2025}.

The recent three-dimensional study referenced in Ref.~\cite{Mataragkas2026} enables us to extend our comparisons beyond two dimensions. Although that study could not apply the field-mixing analysis in the immediate vicinity of the tricritical point (as discussed below), it reports transition fields $\Delta^*_\infty(T)$ along the first-order line of the simple cubic lattice for temperatures ranging from $0.5$ to $1.3$. To apply our procedure to this data, we need to extrapolate the slope over $8\%$ in temperature, from the last tabulated point to the tricritical temperature $T_t = 1.4182(55)$ as reported by Deserno~\cite{Deserno1997}. The local slope, defined as $r(T) = Td\Delta^*/dT - \Delta^*$, increases in magnitude from $-3.01$ at $T \simeq 0.55$ to $-3.20$ at $T \simeq 1.15$, then levels off at the upper end of the data. As a result, the extrapolated value depends on how the curve is extended. Polynomial fits in $T$ and $\beta$ that accurately describe the data within their uncertainties yield values between $-3.06$ and $-3.20$. Additionally, fits in $T$ also extrapolate to $\Delta^*(T_t) = 2.845$-$2.847$, which is consistent with Deserno's value of $\Delta_t = 2.84479(30)$. By taking the center and half-width of this range, we find $r = -3.13(7)$, which translates to $r/\Delta_0 = -1.04(2)$. Notably, using the tricritical point from Deng and Blöte~\cite{Deng2004} alters $r/\Delta_0$ by less than $0.01$.

Additionally, we observe a size dependence in the mixing parameter itself. On both lattices, $s_L$ decreases monotonically with increasing lattice size $L$ at a fixed temperature, but the two cases exhibit qualitative differences at the tricritical point. For the square lattice, we find that $s_L(T_{t}) = 0.093$, $0.048$, $0.035$, and $0.004$ for $L = 12$, $16$, $18$, and $24$, respectively. The zero crossing of $s_L(T)$ shifts from $T \simeq 0.625$ to $T \simeq 0.610$ over this range; therefore, the data are consistent with the conclusion that $s_\infty(T_{t})=0$. In contrast, for the honeycomb lattice, $s_L$ decreases only from $0.388$ to $0.326$ as $L$ increases from $12$ to $24$, showing no signs of approaching zero. Whether the behavior observed in the square lattice indicates a genuine vanishing of the mixing parameter at tricriticality or merely reflects a stronger finite-size trend cannot be determined with the available sizes. This observation is particularly relevant to the practical difficulties discussed in Sec.~\ref{sec:tricritical} and in Ref.~\cite{Mataragkas2026}. If $s$ tends to zero precisely where it is being evaluated, the equal-height condition will lose its discriminatory power in that region, making the numerical solution significantly more fragile.

Table~\ref{tab:ratio} summarizes these determinations. Within the framework of mean-field theory, the Blume–Capel tricritical point is situated at $\mu_{t} = \ln 4$ and $\beta_{t} = \frac{3}{z}$. The second-order line is described by the equation $\beta z = \frac{1}{2} e^{\mu} + 1$. Differentiating at $\mu_{t}$ yields $\frac{{\rm d}\mu}{{\rm d}\beta} = \frac{z}{2}$. Therefore, we obtain the following result:
\begin{align}
  \left(\frac{r}{\Delta_0}\right)_{\rm MF} = -1
  \qquad \text{exactly, for any } z .
  \label{eq:rmf}
\end{align}
This mean-field value is therefore independent of the coordination number. The three two-dimensional lattices, which span $z=3$ to $6$, all give $r/\Delta_0$ between $-1.057$ and $-1.067$: about $6\%$ beyond the mean-field value, with a spread comparable to the systematic uncertainty of a single determination. No dependence on the coordination number is resolved.

The simple-cubic estimate, $r/\Delta_0=-1.04(2)$, lies between the two-dimensional values and the mean-field limit. Compared with the triangular lattice, which has the same coordination number $z=6$, it is closer to $-1$ by $0.03$, a difference of about $1.5$ standard deviations. Since the three-dimensional uncertainty is dominated by the extrapolation to $T_t$, these data cannot establish whether the deviation from Eq.~\eqref{eq:rmf} decreases with increasing dimensionality. We therefore report the near-constancy of $r/\Delta_0$ in two dimensions as an empirical observation, and regard the three-dimensional value as an indication that calls for a determination of the coexistence line closer to the tricritical point.
\begin{table}[tb]
\caption{The field-mixing parameter $r$, normalized by the zero-temperature field $\Delta_0=zJ/2$, across lattices of different coordination numbers and dimensionality. The value for the honeycomb lattice is derived from this study, utilizing Wang–Landau data for lattice sizes $L = 12, 16, 18, 24$. The first of the two square-lattice entries also comes from this work, while the value for the triangular lattice is taken from Ref.~\cite{Mataragkas2025}. It is important to note that the mean-field result presented in Eq.~\eqref{eq:rmf} is exact and independent of the coordination number $z$.}
\footnotetext[1]{From the slope of the multicanonical coexistence line at the tricritical point, Sec.~\ref{sec:muca}.}
\footnotetext[2]{Obtained by applying the present analysis to the transition fields tabulated in Ref.~\cite{Kwak2015} ($L\le48$).} \footnotetext[3]{Obtained by applying Eq.~\eqref{eq:rslope} to the transition fields tabulated in Ref.~\cite{Mataragkas2026} ($0.5\le T\le1.3$), extrapolated to the tricritical point of Ref.~\cite{Deserno1997}; the uncertainty is the spread over polynomial extrapolations.}
\label{tab:ratio}
\begin{ruledtabular}
\begin{tabular}{lccd}
Lattice & $d$ & $z$ & \multicolumn{1}{c}{$r/\Delta_0$} \\
\hline
honeycomb      & 2 & 3 & -1.057(13) \\
honeycomb\footnotemark[1]
               & 2 & 3 & -1.058(3) \\
square         & 2 & 4 & -1.065(12) \\
square\footnotemark[2]
               & 2 & 4 & -1.061(11) \\
triangular     & 2 & 6 & -1.067 \\
simple cubic\footnotemark[3]
               & 3 & 6 & -1.04(2) \\
\hline
mean field     & --- & any & -1 \\
\end{tabular}
\end{ruledtabular}
\end{table}

\subsection{Tricritical eigenvalue exponents}

Near the tricritical point, the distribution of $Q$ follows the equation~\cite{Wilding1996}
\begin{align}
  P_L(Q) = L^{d-y_{t}}\,P_o\!\left[L^{d-y_{t}}(Q-\langle Q\rangle)\right],
  \label{eq:collapse}
\end{align}
where the expected value for the two-dimensional tricritical Ising model is $y_t = 9/5$~\cite{Nijs1979,Nienhuis1979}. To test this equation, we need the absolute scale of $Q$, which requires both mixing parameters.
\begin{figure}[tb]
   \includegraphics[width=\columnwidth]{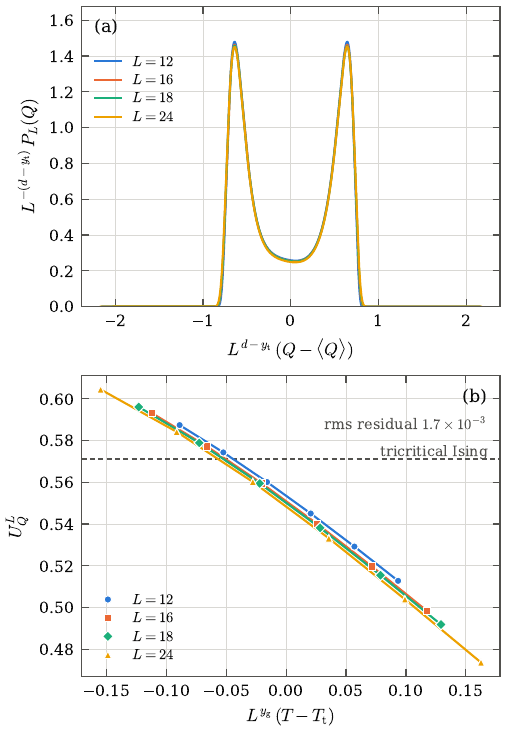}
   \caption{\label{fig:fig3} Finite-size scaling tests of the expected tricritical eigenvalue exponents. (a) Collapse of $P_L(Q)$ with $y_{t}=9/5$. (b) Collapse of $U_Q^L$ with $y_{g}=4/5$.}
\end{figure}

The data for $L = 12$ to $24$ show good agreement, collapsing onto a common curve, as shown in Fig.~\ref{fig:fig3}(a) when using the expected value of $y_t = 9/5$. While this visual agreement serves as a qualitative scaling check, it isn't highly discriminating because $d - y_t = 1/5$ is relatively small. Consequently, the rescaling factor $L^{1/5}$ varies by only 15\% across the available size range. To enhance our analysis, we utilize the fact that Eq.~\eqref{eq:collapse} is equivalent to $\sigma_Q \sim L^{-(d-y_t)}$ and extract the exponent through a log-log fit, which offers a more sensitive evaluation across the available sizes.

The resulting exponent significantly depends on the temperature at which the distributions are analyzed. We obtain values of $y_t = 1.8618$, $1.8170$, $1.7676$, and $1.7160$ at temperatures $T = 0.395$, $0.400$, $0.405$, and $0.410$, respectively, leading to an approximate value of $\frac{{dy_t}}{{dT}} \simeq -9.7$. By interpolating to the independently determined tricritical temperature, we find
\begin{align}
  y_{t} = 1.795(7),
\end{align}
which is consistent with $9/5$ within $0.7\sigma$; the error margin is primarily influenced by the uncertainty surrounding $T_t$. Conversely, if we impose $y_t = 9/5$, we deduce $T = 0.40175$, which is again within $1.1\sigma$ of the cumulant-crossing estimate. It is essential to note that this analysis does not independently determine $y_t$ and $T_t$; rather, it establishes that the pair $(T_t, y_t) = (0.4022, 9/5)$ is self-consistent.

For the second eigenvalue exponent, the cumulant is expected to follow the equation:
\begin{align}
  U_Q^L = U_o\!\left[L^{y_{g}}(T-T_{t})\right],
  \label{eq:collapseUQ}
\end{align}
with $y_{g} = \frac{4}{5}$. The related scaling plot is shown in Fig.~\ref{fig:fig3}(b). When we focus on the range $|T - T_{t}| \le 0.03$, where the scaling form is applicable, the data from four different lattice sizes collapse with a root-mean-square residual of $1.7 \times 10^{-3}$ when plotted against the tricritical scaling variable. A scan over $y_{g}$ reveals a shallow minimum between $0.8$ and $0.9$. This indicates that while the collapse supports the value $y_{g} = \frac{4}{5}$, it does not provide a precise constraint on $y_{g}$ at the $0.1$ level. Collectively, panels (a) and (b) of Fig.~\ref{fig:fig3} offer complementary tests of finite-size scaling for the two expected tricritical eigenvalue exponents, rather than providing independent, high-precision determinations of both exponents.

\section{Interface tension and coexistence line}
\label{sec:muca}

\subsection{Interface tension}

At the pseudo-transition field, the distribution described by Eq.~\eqref{eq:mucarw} is double-peaked. The suppression of the mixed states between the peaks allows us to measure the free energy associated with the two interfaces imposed by the periodic boundary conditions~\cite{Binder1982, Lee1990}. We define $\Delta^*_L$ based on the equal-height condition of $P_L(n)$, and we use the following equation:
\begin{align}
  \Sigma(L) = \frac{1}{2\beta L}\,\ln\frac{P_L^{\max}}{P_L^{\min}} ,
  \label{eq:sigma}
\end{align}
where $P_L^{\min}$ represents the minimum value between the two peaks. The factor of $1/2$ accounts for the presence of both interfaces. It's important to note that this definition of $\Delta^*_L$, which is derived from the distribution of $n$, differs from the field-mixing definition introduced in Sec.~\ref{sec:tricritical} that employs the mixed operator $Q$. However, the two definitions converge as $L \to \infty$.
\begin{figure*}[tb]
   \includegraphics[width=\textwidth]{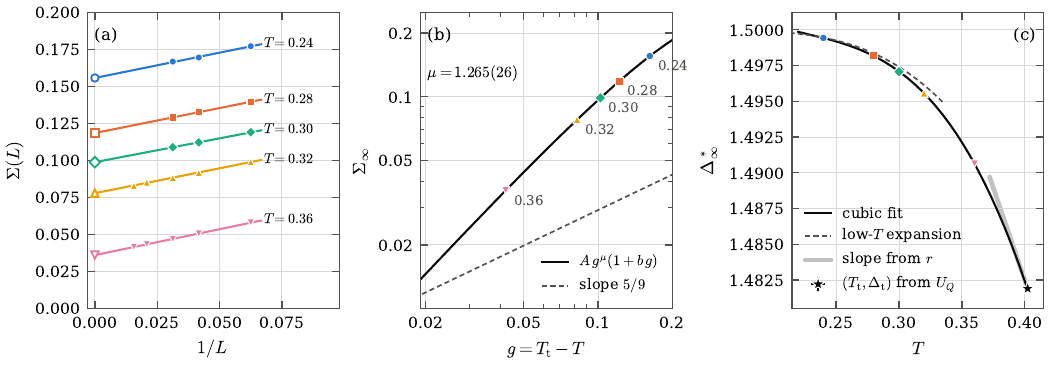}
   \caption{\label{fig:fig4} (a) Size dependence of the interface tension. The open symbols are the extrapolated $\Sigma_\infty$. (b) $\Sigma_\infty$ against the distance $g=T_{t}-T$ from the tricritical point, on logarithmic axes. The solid line is the fit with $\mu$ free and a leading correction. The dashed line shows the slope $5/9$ for comparison. (c) The coexistence line \(\Delta^*_\infty(T)\) is depicted, with a cubic fit shown as a solid line and a low-temperature expansion of Eq.~\eqref{eq:lowT} represented by a dashed line. The slope at the tricritical point is indicated by the field-mixing value of \(r\) through Eq.~\eqref{eq:rslope} (in grey), and the tricritical point is marked by a star, derived from the cumulant analysis. Colors and symbols are used to identify different temperatures in all panels.}
 \end{figure*}
 
The multicanonical results are summarized in Fig.~\ref{fig:fig4}. Fig.~\ref{fig:fig4}(a) displays the finite-size dependence of the interface tension and its extrapolation to the thermodynamic limit. At temperatures $T = 0.32$ and $0.36$, where $L$ ranges from $16$ to $64$, the data is well described by the form:
\begin{align}
  \Sigma(L) = \Sigma_\infty + b\,L^{-1}.
  \label{eq:sigmaL}
\end{align}
This results in $\chi^2/\nu = 9.5/3$ for $T = 0.32$ and $\chi^2/\nu = 1.3/3$ for $T = 0.36$; including an $L^{-2}$ or $L^{-1}\ln L$ term does not significantly reduce $\chi^2$. The larger $\chi^2$ value at $T = 0.32$ indicates an alternation of the residuals between $L = 48$ and $64$ rather than a curvature issue, so we inflate the error accordingly. 

Restricting the fits to $L \leq 32$ reduces $\Sigma_\infty$ by $4 \times 10^{-4}$ and $5 \times 10^{-4}$ at the two temperatures. Since this shift has the same sign in both cases, we add $5 \times 10^{-4}$ in quadrature to the error of the three temperatures for which only $L \leq 32$ data is available. The extrapolated values are provided in Table \ref{tab:muca}.
\begin{table}[tb]
\caption{Multicanonical results along the first-order line: interface tension, from Eq.~\eqref{eq:sigmaL}, and transition field, from the $L^{-2}$ extrapolation of the equal-height condition on $P_L(n)$. For $L\le32$ the error of $\Sigma_\infty$ includes a systematic contribution of $5\times10^{-4}$ (see text).}
\label{tab:muca}
\begin{ruledtabular}
\begin{tabular}{cccc}
$T$ & $L$ & $\Sigma_\infty$ & $\Delta^*_\infty$ \\
\hline
0.24 & 16--32 & 0.1558(13)  & 1.499426(91) \\
0.28 & 16--32 & 0.11864(67) & 1.498174(47) \\
0.30 & 16--32 & 0.09895(84) & 1.497064(79) \\
0.32 & 16--64 & 0.07800(26) & 1.495558(11) \\
0.36 & 16--64 & 0.03616(13) & 1.490630(10) \\
\end{tabular}
\end{ruledtabular}
\end{table}

The interface tension is expected to disappear as we approach the tricritical point from the first-order side. In two dimensions, the relation is given by $\Sigma \sim \xi^{-(d-1)} = \xi^{-1}$. Along the first-order line, the field $\lambda$ goes to zero by construction, and the distance from the tricritical point is measured by $g \propto T_t - T$. Consequently, we find that $\xi \sim g^{-1/y_g}$. Therefore, we have:
\begin{align}
  \Sigma_\infty \propto g^{\mu}, \qquad
  \mu = \frac{d-1}{y_{g}} = \frac{5}{4},
  \label{eq:sigmamu}
\end{align}
for the tricritical Ising class. By fitting $\Sigma_\infty = A\,g^{\mu}(1+b\,g)$, where $g = T_t - T$ and including a leading analytic correction, to the five temperatures listed in Table~\ref{tab:muca}, we obtain:
\begin{align}
  \mu = 1.265 \pm 0.016\,({\rm stat}) \pm 0.020\,(T_{t}),
\end{align}
with the second error arising from shifting $T_t$ by one standard deviation. Figure~\ref{fig:fig4}(b) displays $\Sigma_\infty$ as a function of $g = T_t - T$ on logarithmic axes, clearly illustrating the vanishing of the interface tension as we approach tricriticality. If we impose $\mu = 5/4$, we find $A = 2.026(1)$ and $b = -1.543(4)$ with $\chi^2/\nu = 0.95/3$. In contrast, the dashed reference behavior with exponent $1/y_t = 5/9$, which is associated with the thermal direction across the phase boundary, is evidently incompatible with the data, yielding $\chi^2/\nu = 347/3$ with the same correction term.

The accessible temperature window is limited on both sides. As we move toward $T_t$, the interfacial correlation length, estimated as $\xi \simeq (\beta \Sigma_\infty)^{-1}$, increases from $1.5$ lattice spacings at $T = 0.24$ to 10 at $T = 0.36$. 

On the low-temperature end, several independent runs at $T = 0.20$ and $L = 32$ remained trapped for a significant portion of the simulation, leading to identical weights, but with histograms that differed by factors of $e^{10}$. This behavior indicates barriers orthogonal to $E_2$, which are associated with changes in droplet shape at fixed density. A multicanonical ensemble focused solely on $E_2$ does not eliminate these barriers~\cite{Neuhaus2003}. As a result, we discarded those data. The value $\Sigma_\infty = 0.1893(19)$ obtained from $L = 16$ and $24$ alone is consistent with the extrapolated curve $0.1890$; however, we do not include it in the fit.

\subsection{Coexistence line}

The transition fields outlined in Table~\ref{tab:muca} follow the first-order form described in Eq.~\eqref{eq:fss1st}. At temperatures $T = 0.32$ and $0.36$, the $L^{-2}$ fits over the range of $L = 16$ to $64$ yield results of $\chi^2/\nu = 3.6/3$ and $5.4/3$, respectively. In contrast, an $L^{-1}$ form is rejected, showing values of $\chi^2/\nu = 29/3$ and $104/3$. This extends the finite-size consistency check discussed in Sec.~\ref{sec:tricritical} to sizes much larger than $L = 24$.

At $T = 0.36$, the multicanonical estimate of $1.490630(10)$ is $3 \times 10^{-5}$ below the WL value listed in Table~\ref{tab:firstorder}. When restricting the multicanonical data to $L \leq 32$, we obtain $1.490644(51)$, which is in full agreement with the previous results. Thus, the small difference reflects the influence of larger sizes rather than the method used.

Figure~\ref{fig:fig4}(c) presents the thermodynamic-limit coexistence fields $\Delta_\infty^*(T)$ and compares them with the tricritical endpoint, the local slope implied by field mixing, and the low-temperature expansion. These comparisons provide three complementary checks that connect the multicanonical coexistence line to the preceding results.

\textit{Endpoint:} The coexistence line corresponds to the locus $\lambda = 0$ of an analytic scaling field; therefore, it is analytic at $T_{t}$, where it smoothly transitions into the critical line. A polynomial in $T$ is an appropriate representation. A quadratic fit is rejected ($\chi^2/\nu = 35/2$), while a cubic fit is acceptable ($\chi^2/\nu = 0.2/1$), and it extrapolates to $\Delta^*(T_{t}) = 1.48185(23)$, which is consistent with $\Delta_{t} = 1.4820(2)$ obtained from the cumulant analysis.

\textit{Slope:} Along the line where $\lambda=0$, we have $\frac{{\rm d}\mu}{{\rm d}\beta} = -r$. By substituting $\mu = \beta \Delta$, this relationship can be expressed as
\begin{align}
  r = T\,\frac{{\rm d}\Delta^*}{{\rm d}T} - \Delta^*\Big|_{T=T_{t}}.
  \label{eq:rslope}
\end{align}
Differentiating the cubic function at $T_t$ yields
\begin{align}
  r = -1.587 \pm 0.004\,({\rm stat}) \pm 0.002\,({\rm syst}),
\end{align}
where the systematic error accounts for the uncertainty in $T_t$ as well as the variability derived from anchoring the polynomial at $(T_t, \Delta_t)$. This determination does not utilize the operator $Q$ or the mixing parameter $s$, making it independent of the field-mixing value $r = -1.585(19)$ mentioned in Sec.~\ref{sec:exponents}. Notably, it agrees with this value while being four times more precise. This result corresponds to $r/\Delta_0 = -1.058(3)$ (see Table~\ref{tab:ratio}).

\textit{Low Temperature:} At $\Delta = \Delta_0 = zJ/2$, a connected cluster of $k$ vacancies in the ferromagnetic phase and a connected cluster of $k$ parallel spins in the phase dominated by $\sigma = 0$ have the same energy, which is $(z/2)k$ minus the number of bonds within the cluster. However, the latter cluster can be oriented in two ways. The additional entropy favors the $\sigma = 0$ phase, causing the coexistence line to bend below $\Delta_0$.

To second order in the cluster expansion (considering isolated sites and the $zN/2$ nearest-neighbor pairs), the expression is given by:
\begin{align}
  \Delta^*(T) = \frac{z}{2} - T\left[e^{-z/2T} + \frac{z}{2}\,e^{-(z-1)/T}\right] + \cdots
  \label{eq:lowT}
\end{align}
At $T = 0.24$, this equation agrees with the multicanonical value to within $0.3 \sigma$. At $T = 0.28$ and $0.30$, the deviation decreases from $11\sigma$ to $4\sigma$ when comparing first and second order approximations. This indicates that the second-order truncation brings the expansion closer to the simulation data. However, as one approaches $T_{t}$, this expansion breaks down, as expected.

\section{Second-order line and phase diagram}

The estimated joint density of states remains temperature-independent, allowing us to use the same WL data to map the continuous branch of the phase boundary. For each value of $\Delta < \Delta_{t}$, we identify the maximum of the magnetic susceptibility as a function of temperature $T$ for each lattice size and perform an extrapolation according to the following equation:
\begin{align}
  T_{c}(L) = T_{c} + a\,L^{-1/\nu},
\end{align}
where the exact two-dimensional Ising exponent is $\nu = 1$. The resulting finite-size estimates are presented in Table~\ref{tab:secondorder}. At zero crystal field, we find $T_{c}(\Delta=0) = 1.157$. This value can be compared with the effective-field estimates of $1.3515$ (CEFT) and $1.5192$ (EFT)~\cite{Kaneyoshi1990}. The ratio of our result to the exact spin-$\frac{1}{2}$ Ising value, $T_{c}^{\text{Ising}} = \frac{2}{\ln(2+\sqrt{3})} \approx 1.5187$, is $0.762$, which is close to the previously reported values of $0.746$ and $0.735$ found on square and triangular lattices.
\begin{table}[tb]
\caption{Critical temperature along the second-order branch, from the finite-size extrapolation of the susceptibility maxima using $\nu=1$. Values with $\Delta \gtrsim 1.2$ should be interpreted with caution, as near the tricritical point, the appropriate exponent transitions from the Ising value $\nu = 1$ to the tricritical value $\nu_{t} = \frac{1}{y_{t}} = \frac{5}{9}$. This transition causes the extrapolation to be systematically biased in that region.}
\label{tab:secondorder}
\begin{ruledtabular}
\begin{tabular}{dd}
\multicolumn{1}{c}{$\Delta$} & \multicolumn{1}{c}{$T_{c}$} \\
\hline
0.00 & 1.1571 \\
0.25 & 1.1054 \\
0.50 & 1.0350 \\
0.75 & 0.9564 \\
1.00 & 0.8489 \\
1.20 & 0.7407 \\
1.35 & 0.6223 \\
1.45 & 0.4914 \\
\end{tabular}
\end{ruledtabular}
\end{table}

\begin{figure*}[tb]
   \includegraphics[width=\textwidth]{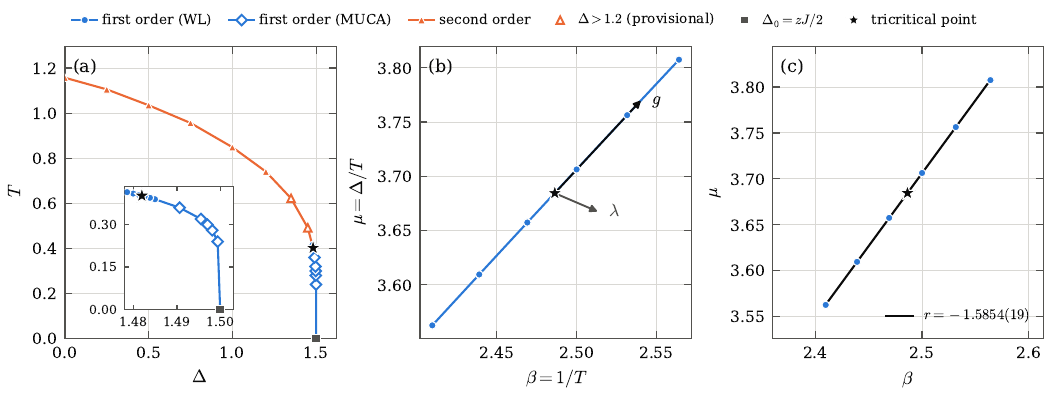}
   \caption{\label{fig:fig5} (a) Phase diagram of the honeycomb Blume--Capel ferromagnet in the $(\Delta,T)$ plane. (b) The same boundary in the $\beta$-$\mu$ plane, with the directions of the scaling fields $g$ and $\lambda$. (c) Enlargement near the tricritical point; the slope gives $r=-1.585(2)_{\rm stat}$ before inclusion of the systematic uncertainty discussed in the text. The linear fit lies within $|T-T_t|\le0.015$.}
\end{figure*}

Figure~\ref{fig:fig5} summarizes the results of the phase boundary obtained from various numerical analyses. Fig.~\ref{fig:fig5}(a) displays the phase diagram in the $(\Delta,T)$ plane, illustrating how the continuous and first-order branches intersect at the tricritical point. The first-order branch approaches the zero-temperature endpoint, given by $(\Delta_0,0)=(3/2,0)$. Fig.~\ref{fig:fig5}(b) shows the same boundary in the $\beta$-$\mu$ plane, where the local directions of the scaling fields $g$ and $\lambda$ clarify the field-mixing construction. Fig.~\ref{fig:fig5}(c) provides an enlarged view of the tricritical region in this representation; the local slope of the coexistence line defines the mixing parameter $r$. The consistent results for the phase boundaries from WL, HR, and MUCA analyses present a unified view of the two branches and their tricritical junction.

Outside the tricritical region, the maxima of the susceptibility align with the Ising finite-size scaling form. However, as $\Delta$ approaches $\Delta_{t}$, the extrapolation becomes increasingly unstable, indicating a crossover from Ising to tricritical scaling. This change corresponds to the correlation-length exponent shifting from $\nu=1$ to $\nu_{t}=1/y_{t}=5/9$, meaning that no single power law can accurately describe the shift of the maxima in the crossover region. Consequently, we treat the estimates for $\Delta \gtrsim 1.2$ as provisional and do not use them to independently determine the tricritical point. The location of $(\Delta_{t},T_{t})$, mentioned in Sec.~\ref{sec:tricritical}, relies entirely on the field-mixing analysis. A quantitative approach to the crossover region will necessitate larger lattices and a scaling form that interpolates between the Ising and tricritical fixed points.

\section{Summary and outlook}

We have mapped the phase boundary of the spin-1 Blume-Capel ferromagnet on the honeycomb lattice and identified the tricritical point at $(\Delta_{t}, T_{t})=[1.4820(2),0.4022(7)]$. This result is supported by two independent simulation methods: Wang-Landau sampling of the joint density of states and histogram reweighting from Metropolis runs at larger lattice sizes, both of which yield mutually consistent crossing estimates. Additionally, we obtained the field-mixing parameter $r=-1.585(19)$. We demonstrated that the combination $r/\Delta_0$ equals $-1$ in mean-field theory for any coordination number. We collected corresponding values for four lattices, three of which had not been previously determined. The three two-dimensional lattices, from $z=3$ to $z=6$, cluster around $r/\Delta_0\simeq-1.06$, while the simple-cubic value, $-1.04(2)$, lies between this cluster and mean field, with an uncertainty dominated by the extrapolation to the tricritical temperature. Our findings showed finite-size scaling consistent with the expected tricritical Ising exponents $y_t=9/5$ and $y_g=4/5$ across the available lattice sizes.

Multicanonical simulations up to $L=64$, using weights constructed directly from the Wang-Landau density-of-states estimates for $L\le24$ and refined for larger sizes, allowed us to access the interface tension along the first-order line. The approach of the interface tension to zero is described by $(T_t-T)^{\mu}$ with $\mu=1.265(26)$, which is consistent with the tricritical Ising value of $1/y_{g}=5/4$ but incompatible with $1/y_{t}=5/9$. The extrapolation of the coexistence line aligns with the tricritical point obtained from the cumulant analysis, and its slope provides an independent determination of the mixing parameter, $r=-1.587(5)$, which is four times more precise than the initial field-mixing estimate.

Several questions remain unresolved. The critical exponents along the second-order branch, the anticipated logarithmic divergence of the specific heat, and the immediate vicinity of the tricritical point—where the Ising-based extrapolations from Table~\ref{tab:secondorder} become unreliable—require significantly larger lattices. At $T=0.38$, the condition $L\gtrsim6\xi$ already necessitates $L\approx150$. At the low-temperature end, a multicanonical ensemble in $E_2$ alone becomes inefficient near $T=0.20$, where tunneling between the phases is constrained by barriers orthogonal to $E_2$~\cite{Neuhaus2003}; thus, a second reaction coordinate would be necessary in that regime.

We also note two features of the current data that warrant further investigation. First, the mixing parameter $s_L$ obtained from the histogram-reweighting process is significantly less stable than that from the Wang-Landau data. This instability reflects the weak sensitivity of the equal-height condition to $s$ at large $L$; therefore, the reweighting data were excluded from the extrapolation of Eq.~\eqref{eq:fss1st}. Second, the value of $U_Q$ at successive crossings drifts gradually away from the expected universal value of $0.571$ as $L$ increases, moving from $0.563$ at $12\times16$ to $0.534$ at $36\times48$. Whether this reflects genuine corrections to scaling or the aforementioned instability in $s_L$ remains to be determined by future data.

These observations can be interpreted in light of the three-dimensional study reported in Ref.~\cite{Mataragkas2026}, which indicates that the field-mixing procedure ceases to converge when the two peaks of $P_L(\tilde{Q})$ become only weakly separated. It identifies an empirical threshold $U_Q \lesssim 0.3$ beyond which the equal-height and equal-population conditions lose their discriminatory power. A similar mechanism may be relevant here: as $s_L$ decreases with increasing $L$, the equal-height condition becomes progressively less sensitive to $s$, with its numerical solution increasingly governed by the discretization of the joint density of states rather than by the physics of the system. We note that in the present two-dimensional calculation, the cumulant remains comfortably above that threshold throughout, the smallest value encountered during our analysis is $U_Q=0.47$, occurring at $L=24$ and $T=0.415$. This suggests that the honeycomb data remain within the empirically stable regime of the method.

\begin{acknowledgments}
The authors thank LaMCAD/UFG for the computational resources.
\end{acknowledgments}

\bibliography{references}

@article{Blume1966,
  author  = {Blume, M.},
  title   = {Theory of the First-Order Magnetic Phase Change in {UO$_2$}},
  journal = {Phys. Rev.},
  volume  = {141},
  pages   = {517},
  year    = {1966},
  doi     = {10.1103/PhysRev.141.517}
}

@article{Capel1966,
  author  = {Capel, H. W.},
  title   = {On the possibility of first-order phase transitions in {Ising}
             systems of triplet ions with zero-field splitting},
  journal = {Physica},
  volume  = {32},
  pages   = {966},
  year    = {1966}
}

@article{Blume1971,
  author  = {Blume, M. and Emery, V. J. and Griffiths, R. B.},
  title   = {Ising Model for the $\lambda$ Transition and Phase Separation in {H}e$^3$-{H}e$^4$ Mixtures},
  journal = {Physical Review A},
  volume  = {4},
  number  = {3},
  pages   = {1071--1082},
  year    = {1971},
  doi     = {10.1103/PhysRevA.4.1071}
}

@book{DombLebowitz,
  editor    = {Domb, C. and Lebowitz, J. L.},
  title     = {Phase Transitions and Critical Phenomena},
  publisher = {Academic Press},
  address   = {London},
  year      = {1984},
  volume      = {9}
}

@article{Silva2006,
  author  = {Silva, C. J. and Caparica, A. A. and Plascak, J. A.},
  title   = {{Wang-Landau} {Monte Carlo} simulation of the {Blume-Capel}
             model},
  journal = {Phys. Rev. E},
  volume  = {73},
  pages   = {036702},
  year    = {2006},
  doi     = {10.1103/PhysRevE.73.036702}
}

@article{Kwak2015,
  author  = {Kwak, W. and Jeong, J. and Lee, J. and Kim, D.-H.},
  title   = {First-order phase transition and tricritical scaling behavior
             of the {Blume-Capel} model: A {Wang-Landau} sampling approach},
  journal = {Phys. Rev. E},
  volume  = {92},
  pages   = {022134},
  year    = {2015},
  doi     = {10.1103/PhysRevE.92.022134}
}

@article{Zierenberg2017,
  author  = {Zierenberg, J. and Fytas, N. G. and Weigel, M. and
             Janke, W. and Malakis, A.},
  title   = {Scaling and universality in the phase diagram of the {2D}
             {Blume-Capel} model},
  journal = {Eur. Phys. J. Spec. Top.},
  volume  = {226},
  pages   = {789},
  year    = {2017}
}

@article{Mataragkas2025,
  author  = {Mataragkas, D. and Vasilopoulos, A. and Fytas, N. G. and
             Kim, D.-H.},
  title   = {Tricriticality and finite-size scaling in the triangular
             {Blume-Capel} ferromagnet},
  journal = {Phys. Rev. Research},
  volume  = {7},
  pages   = {013214},
  year    = {2025},
  doi     = {10.1103/PhysRevResearch.7.013214}
}

@article{Kaneyoshi1990,
  author  = {Kaneyoshi, T.},
  title   = {Correlated-effective-field treatment of spin-one {Ising}
             models},
  journal = {Physica A},
  volume  = {164},
  pages   = {730},
  year    = {1990}
}

@article{Wilding1992,
  author  = {Wilding, N. B. and Bruce, A. D.},
  title   = {Density fluctuations and field mixing in the critical fluid},
  journal = {J. Phys.: Condens. Matter},
  volume  = {4},
  pages   = {3087},
  year    = {1992}
}

@article{Bruce1992,
  author  = {Bruce, A. D. and Wilding, N. B.},
  title   = {Scaling fields and universality of the liquid-gas critical
             point},
  journal = {Phys. Rev. Lett.},
  volume  = {68},
  pages   = {193},
  year    = {1992},
  doi     = {10.1103/PhysRevLett.68.193}
}

@article{Wilding1996,
  author  = {Wilding, N. B. and Nielaba, P.},
  title   = {Tricritical universality in a two-dimensional spin fluid},
  journal = {Phys. Rev. E},
  volume  = {53},
  pages   = {926},
  year    = {1996},
  doi     = {10.1103/PhysRevE.53.926}
}

@article{Ferrenberg1988,
  author  = {Ferrenberg, A. M. and Swendsen, R. H.},
  title   = {New {Monte Carlo} technique for studying phase transitions},
  journal = {Phys. Rev. Lett.},
  volume  = {61},
  pages   = {2635},
  year    = {1988},
  doi     = {10.1103/PhysRevLett.61.2635}
}

@article{Nijs1979,
  author  = {den Nijs, M. P. M.},
  title   = {A relation between the temperature exponents of the
             eight-vertex and $q$-state {Potts} model},
  journal = {J. Phys. A: Math. Gen.},
  volume  = {12},
  pages   = {1857},
  year    = {1979}
}

@article{Nienhuis1979,
  author  = {Nienhuis, B. and Berker, A. N. and Riedel, E. K. and
             Schick, M.},
  title   = {First- and Second-Order Phase Transitions in {Potts} Models:
             Renormalization-Group Solution},
  journal = {Phys. Rev. Lett.},
  volume  = {43},
  pages   = {737},
  year    = {1979},
  doi     = {10.1103/PhysRevLett.43.737}
}

@article{Berg1992,
  author  = {Berg, B. A. and Neuhaus, T.},
  title   = {Multicanonical ensemble: A new approach to simulate
             first-order phase transitions},
  journal = {Phys. Rev. Lett.},
  volume  = {68},
  pages   = {9},
  year    = {1992},
  doi     = {10.1103/PhysRevLett.68.9}
}

@article{Lee1990,
  author  = {Lee, J. and Kosterlitz, J. M.},
  title   = {New numerical method to study phase transitions},
  journal = {Phys. Rev. Lett.},
  volume  = {65},
  pages   = {137},
  year    = {1990},
  doi     = {10.1103/PhysRevLett.65.137}
}

@article{Mataragkas2026,
title = {First-order phase transitions in the three-dimensional {Blume–Capel} ferromagnet},
journal = {Physica A: Statistical Mechanics and its Applications},
volume = {699},
pages = {131922},
year = {2026},
issn = {0378-4371},
author = {D. Mataragkas and A. Vasilopoulos and Z.D. Vatansever and Y. Kim and D.-H. Kim and N.G. Fytas}
}

@article{Deserno1997,
  author  = {Deserno, M.},
  title   = {Tricriticality and the {Blume-Capel} model: A {Monte Carlo}
             study within the microcanonical ensemble},
  journal = {Phys. Rev. E},
  volume  = {56},
  pages   = {5204},
  year    = {1997},
  doi     = {10.1103/PhysRevE.56.5204}
}

@article{Deng2004,
  author  = {Deng, Youjin and Bl\"ote, Henk W. J.},
  title   = {Constrained tricritical {Blume-Capel} model in three
             dimensions},
  journal = {Phys. Rev. E},
  volume  = {70},
  pages   = {046111},
  year    = {2004},
  doi     = {10.1103/PhysRevE.70.046111}
}

@article{Binder1982,
  author  = {Binder, K.},
  title   = {{Monte Carlo} calculation of the surface tension for two- and
             three-dimensional lattice-gas models},
  journal = {Phys. Rev. A},
  volume  = {25},
  pages   = {1699},
  year    = {1982},
  doi     = {10.1103/PhysRevA.25.1699}
}

@article{Neuhaus2003,
  author  = {Neuhaus, T. and Hager, J. S.},
  title   = {{2D} crystal shapes, droplet condensation, and exponential
             slowing down in simulations of first-order phase transitions},
  journal = {J. Stat. Phys.},
  volume  = {113},
  pages   = {47--83},
  year    = {2003},
  doi     = {10.1023/A:1025718703965}
}

@article{Ibagon2021,
  author    = {Ibagon, Ingrid and Furlan, A. P. and Oliveira, T. J. and Dickman, Ronald},
  title     = {Phase diagram and critical properties of a two-dimensional associating lattice gas},
  journal   = {Physical Review E},
  volume    = {104},
  number    = {6},
  pages     = {064120},
  year      = {2021},
  publisher = {American Physical Society},
}

@Article{Caparica2014,
  Title                    = {{W}ang-{L}andau sampling: A criterion for halting the simulations},
  Author                   = {Caparica, A. A.},
  Journal                  = {Phys. Rev. E},
  Year                     = {2014},
  Month                    = {Apr},
  Pages                    = {043301},
  Volume                   = {89},
  Issue                    = {4},
  Numpages                 = {7}
}

@Article{Caparica2012,
  Title                    = {{W}ang-{L}andau sampling: {I}mproving accuracy},
  Author                   = {Caparica, A. A. and Cunha-Netto, A. G.},
  Journal                  = {Phys. Rev. E},
  Year                     = {2012},
  Month                    = {Apr},
  Pages                    = {046702},
  Volume                   = {85},
  Issue                    = {4},
  Numpages                 = {8}
}

@article{Ferreira2018,
  title={Wang--Landau sampling: Saving {CPU} time},
  author={Ferreira, LS and Jorge, LN and Le{\~a}o, SA and Caparica, AA},
  journal={Journal of Computational Physics},
  volume={358},
  pages={130--134},
  year={2018},
  publisher={Elsevier}
}

@Article{Wang2001,
  Title                    = {Determining the density of states for classical statistical models: A random walk algorithm to produce a flat histogram},
  Author                   = {Wang, Fugao and Landau, DP},
  Journal                  = {Physical Review E},
  Year                     = {2001},
  Number                   = {5},
  Pages                    = {056101},
  Volume                   = {64}
}

@Article{Wang2001a,
  Title                    = {Efficient, multiple-range random walk algorithm to calculate the density of states},
  Author                   = {Wang, Fugao and Landau, DP},
  Journal                  = {Physical review letters},
  Year                     = {2001},
  Number                   = {10},
  Pages                    = {2050},
  Volume                   = {86}
}

@Book{Landau2014,
  Title                    = {A guide to {M}onte {C}arlo simulations in statistical physics},
  Author                   = {Landau, David P and Binder, Kurt},
  Publisher                = {Cambridge University Press},
  Year                     = {2014}
}

@article{Booth1993,
title = {{Monte Carlo} study of the BEG model on a honeycomb lattice},
journal = {Journal of Magnetism and Magnetic Materials},
volume = {128},
number = {1},
pages = {117-123},
year = {1993},
issn = {0304-8853},
author = {R.J.C. Booth and Lu Hua and J.W. Tucker and C.M. Care and I. Halliday}
}

\end{document}